\documentclass[runningheads]{llncs}
\usepackage[table]{xcolor}
\usepackage[utf8]{inputenc}
\usepackage{url}
\usepackage[hyperfootnotes=false]{hyperref}
\usepackage{amssymb}
\usepackage{wasysym}
\usepackage{rotating}
\usepackage{lscape}
\usepackage{graphicx}
\usepackage{hhline}
\usepackage{multirow,makecell}
\usepackage{boldline} 
\usepackage{tikz}
\usetikzlibrary{patterns}
\usepackage{array}
\usepackage{ifthen}
\usepackage{tablefootnote}
\usepackage{multicol}
\usepackage{footmisc}

\usepackage{siunitx}
\usepackage{booktabs}

\makeatletter
\renewcommand\@makefntext[1]{%
\setlength\parindent{1em}%
\noindent
\mbox{\@thefnmark.~}{#1}}

\makeatother

\newcommand{\graycell}[1]{%
  \if\relax\detokenize{#1}\relax\cellcolor{gray!25}\else$\;\,$\cellcolor{gray!25}\fi
}
\newcommand{\darkgraycell}[1]{%
  \if\relax\detokenize{#1}\relax\cellcolor{gray!80}\else$\;\,$\cellcolor{gray!80}\fi
}
\newcolumntype{x}[1]{%
>{\centering\hspace{0pt}}p{#1}}%

\newcommand{\tn}{\tabularnewline}

\newcolumntype{P}[1]{>{\centering\arraybackslash}p{#1}}

\definecolor{backColor}{RGB}{200,200,200}% grey

\newlength{\hatchspread}
\newlength{\hatchthickness}
\newlength{\hatchshift}
\newcommand{\hatchcolor}{}
\tikzset{hatchspread/.code={\setlength{\hatchspread}{#1}},
         hatchthickness/.code={\setlength{\hatchthickness}{#1}},
         hatchshift/.code={\setlength{\hatchshift}{#1}},% must be >= 0
         hatchcolor/.code={\renewcommand{\hatchcolor}{#1}}}
\tikzset{hatchspread=3pt,
         hatchthickness=0.4pt,
         hatchshift=0pt,% must be >= 0
         hatchcolor=black}

\pgfdeclarepatternformonly[\hatchspread,\hatchthickness,\hatchshift,\hatchcolor]% variables
   {custom north east lines}% name
   {\pgfqpoint{\dimexpr-2\hatchthickness}{\dimexpr-2\hatchthickness}}% lower left corner
   {\pgfqpoint{\dimexpr\hatchspread+2\hatchthickness}{\dimexpr\hatchspread+2\hatchthickness}}% upper right corner
   {\pgfqpoint{\dimexpr\hatchspread}{\dimexpr\hatchspread}}% tile size
   {% shape description
    \pgfsetlinewidth{\hatchthickness}
    \pgfpathmoveto{\pgfqpoint{\dimexpr\hatchshift-0.15pt}{-0.15pt}}
    \pgfpathlineto{\pgfqpoint{\dimexpr\hatchspread+0.15pt}{\dimexpr\hatchspread-\hatchshift+0.15pt}}
    \ifdim \hatchshift > 0pt
      \pgfpathmoveto{\pgfqpoint{-0.15pt}{\dimexpr\hatchspread-\hatchshift-0.15pt}}
      \pgfpathlineto{\pgfqpoint{\dimexpr\hatchshift+0.15pt}{\dimexpr\hatchspread+0.15pt}}
    \fi
    \pgfsetstrokecolor{\hatchcolor}
    \pgfusepath{stroke}
   }

\title{Trust assumptions in voting systems}

\author{Kristjan Krips\inst{1,2}\and Nikita Snetkov\inst{1,3}\orcidID{0000-0002-1414-2080} \and Jelizaveta Vakarjuk\inst{1,3}\orcidID{0000-0001-6398-3663} \and Jan Willemson\inst{1}\orcidID{0000-0002-6290-2099}\\
\email{\{kristjan.krips,nikita.snetkov,jelizaveta.vakarjuk,jan.willemson\}@cyber.ee}}
\institute{Cybernetica AS, Mäealuse 2/1, 12618 Tallinn, Estonia \and
Institute of Computer Science, University of Tartu,
Narva mnt 18, Tartu, Estonia \and Tallinn University of Technology, Akadeemia tee 15a, 12618 Tallinn, Estonia}

\authorrunning{K. Krips, N. Snetkov, J. Vakarjuk, J. Willemson}

\begin{document}
\maketitle

\begin{abstract}
Assessing and comparing the security level of different voting systems is non-trivial as the technical means provided for and societal assumptions made about various systems differ significantly. However, trust assumptions concerning the involved parties are present for all voting systems and can be used as a basis for comparison. This paper discusses eight concrete voting systems with different properties, 12 types of parties involved, and seven general security goals set for voting. The emerging trust relations are assessed for their criticality, and the result is used for comparison of the considered systems.
\end{abstract}

\section{Introduction}
Voting is a widely used approach to determining social preferences. In the case of political elections, voting results may have significant consequences for the given society or even globally. Thus, it is essential to ensure that the result adequately represents the true preferences of society.

As the stakes are high, there are also incentives to manipulate the system. Unfortunately, it is not easy to develop a voting method that would be provably immune to all conceivable problems. 

On the one hand, it would be favourable if every member of the society would be able to certify the correctness of the tally. However, such a goal implies that everyone should know how everyone else voted. This kind of \emph{viva voce} voting has been used throughout the history of democracy, but it unfortunately also has a downside of enabling vote-buying~\cite{Brent2006,Wasley2016}. To mitigate this threat, secret ballots have been used already in ancient Greece~\cite{10.2307/147360}, with our current implementation originating in mid-19th century Australia~\cite{Brent2006}. 

As the usage of secret ballots ensures that votes are anonymous, tally integrity has to be provided with organisational measures like voting in a controlled environment and distributed counting. However, such measures introduce additional trust assumptions. For example, to accept the results of large-scale paper ballot elections, one must assume that the people performing counts and recounts on a subset of ballots were not all corrupt in a coordinated manner. 

Electronic voting has the potential to make the tally computationally verifiable. This functionality, however, comes at the cost of additional protocol complexity and the requirement to be able to verify cryptographic operations. This capability cannot be assumed from the general public, who instead has to rely on the statements made by a few technical experts.

One way or another, trusting some components or actors to a certain extent seems unavoidable for all the currently known voting systems. For example, many of the mitigation measures used by voting schemes rely on election observers and auditors. However, the election organiser has the power to influence how the elections are observed and audited, which means that in some cases, a malicious election organiser may be able to evade audits. Therefore, it is essential to identify the underlying trust assumptions to determine the limitations of election audits.

This paper aims to identify and classify the main trust assumptions across different voting methods. 

To handle this task methodically, we first identified four general classes of voting systems (paper voting in a polling station, postal voting, machine voting, and remote electronic voting). For each of these classes, we also consider subclasses based on the verifiability options that the systems in that class may provide. 

Next, we identified the main parties involved in the respective voting systems. Several parties are universal across various voting methods (voter, registrar, election organiser, infrastructure provider, election observer), some are specific to paper-based voting (polling station official, printing house, postal service), and some to machine voting and electronic voting (voter's computer, hardware vendor, software vendor).

Finally, we established the security requirements of interest in the context of our research. The list of such requirements is not completely universal and varies to an extent from source to source (see e.g.~\cite{DBLP:conf/sec/MitrouGK02,DBLP:conf/IEEEares/Cetinkaya08,DBLP:conf/hicss/Schryen04,heiberg2014modeling}).
However, for the purposes of this paper, we consider the following requirements:
\begin{itemize}
    \item \emph{Ballot secrecy}: It should not be possible to link the content of the vote to the identity of the voter.
    \item \emph{Coercion resistance}: The voter should be able to cast a vote that accurately represents their genuine choice even if coercive agents are present during the voting period.
    \item \emph{Universal suffrage}: Every eligible voter is allowed to participate in the elections.
    \item \emph{Equal suffrage}: The vote of every eligible voter should have equal weight.
    \item \emph{Verification of eligibility}: It should be possible to detect if some votes have been cast by ineligible voters. It should not be possible to undetectably cast a vote on behalf of an eligible voter who has not cast a ballot.
    \item \emph{Verification of delivery}: Upon casting a ballot, a voter should be able to verify that their intent has been properly interpreted and that their vote has been recorded without alteration.
    \item \emph{Verification of ballot box integrity}: It should be impossible to undetectably modify a ballot, remove it from a ballot box or add ineligible votes.
    \item \emph{Verification of tally integrity}: The final election results can be verified as soon as they are announced by the election authorities.
\end{itemize}

With the categories of interest established, we identified which voting systems rely on the trust in certain parties in order to achieve the respective security properties. The results are summarised in Table~\ref{tab:ivoting_threats_corrupted_parties_new}. In order to reduce the subjectivity of this assessment, the table was first composed by one author and then reviewed and improved by the others. The process involved two rounds of discussions between all the authors, going through every cell of the table.

After having composed a systematic overview of the trust assumptions used by different voting systems, we can take one further step and compare the security level of these systems.  Despite a number of previous attempts made in the literature~\cite{DBLP:conf/egov/KrimmerV05,6911807,Neumann,DBLP:conf/sec/NeumannNV17,DBLP:journals/istr/Willemson18,finFramework}, such a comparison still remains largely an open problem. The main reason behind the hardness of this task is that voting security has many aspects. The comparison would need to happen in many dimensions, and it is unlikely that one system would dominate another in all of them. 

However, we will make an attempt in this paper to unify the treatment of different dimensions in terms of trust assumptions. Intuitively, we can call one system more secure compared to another if it relies on a smaller set of assumptions. Interpreting ``smaller'' as a set inclusion, we would still get incomparability, but there are other possible interpretations that we will be presenting in the concluding part of this paper.

\section{Paper-based voting in a polling station}
We will first consider the trust assumptions of regular paper-based voting and then review whether some of the trust assumptions can be removed by providing voters with cryptographic receipts allowing voters to verify their votes.

\subsection{Regular paper voting}
\label{paper_uk}
Traditional paper-based voting systems rely on the distribution of trust to reduce the risk of fraud. However, some trust assumptions remain. For example, the election organiser has to ensure that elections can be observed, and is responsible for organising the printing and delivery of paper ballots and voter lists.

We illustrate the trust assumptions of paper-based voting by examining the voting system used in the United Kingdom. The United Kingdom has a long tradition of electing representatives, and its legal and political system has been taken as a basis in several Commonwealth countries. 

Over time, UK's voting system has had to adapt to the changes in society. For example, as a significant change, the requirement of ballot secrecy was introduced in 1872 as a part of electoral reforms~\cite{CrCr07}. To prevent ballot secrecy from being abused and to allow fraud allegations to be investigated, serial numbers were added to ballots and linked to voters when ballots were picked up. It turns out that 150 years later, a similar paper-based voting system is still in use. 

\textbf{Ballot secrecy}
While polling booths protect voters from prying eyes, it does not guarantee that no recording devices have been hidden on the premises of the polling station. The polling station official who hands out ballots has to be trusted not to link the voter to the serial number on the ballot\footnote{Before handing out the ballot, the election official perforates it, and either writes its serial number into the voter list or voter's name on the ballot's counterfoil~\cite{OSCEUK2005finalreport,OSCEUK2010finalreport}.}. The election court and, in the case of Parliamentary elections, the House of Commons have the right to require investigation of fraud allegations, which may lead to ballot secrecy being violated for the ballots/voters under investigation\footnote{Since 1886, seven investigations have resulted in ballot secrecy being violated~\cite{lawcommissionmannerofvotinguk}.}. To increase the level of ballot secrecy, the ballot boxes are transported to the counting centre of the constituency, where the ballots are mixed and counted~\cite{OSCEUK2005finalreport}.

\textbf{Coercion-resistance} Voters have to be trusted not to use recording devices to take photos or videos of the filled-in ballot, including the ballot's serial number. The polling station official who hands out ballots must be trusted as that person could link the voter to the serial code on the ballot. A coercer who forces voters to abstain might be able to check whether the voter registered to vote\footnote{British Library keeps full versions of electoral registries. The records less than ten years old can be viewed under supervision for research purposes~\cite{britishlibraryelectoralregisters}.}.

\textbf{Universal suffrage}
The voter has to be in an electoral register to be allowed to vote. Thus, the managers of electoral registers are trusted to guarantee the integrity of the registry\footnote{There are 381 electoral registers in the UK~\cite{OSCEUK2017needsassessment}.}. Since 2015, voters can register to vote via a web page or by filling in a paper form\footnote{The registration process is described at \url{https://www.gov.uk/register-to-vote}.}. Thus, the registration web page must be considered a trusted part of the voting system. Election observers can not monitor the registration phase as monitoring is regulated only during election day~\cite{OSCEUK2017needsassessment}.

\textbf{Equal suffrage}
If a voter learns that a vote has already been cast under their name, they are provided with an opportunity to cast a tendered vote, which will only be counted if decided so by the election court~\cite{OSCEUK2005finalreport}. As polling station officials have access to the voter list, they must be trusted not to stuff the ballot box, while the election observers must be trusted to report any misconduct.

\textbf{Verification of eligibility}
Voters are trusted to use valid information when registering to vote. To register, they have to provide their date of birth, national insurance number, and, depending on the voter's residence, either a signature or an address\footnote{Since 2002, Northern Ireland has required a signature, date of birth, and national insurance number to register to vote~\cite{OSCEUK2010finalreport}. After individual electoral registration (IER) was introduced in 2014, voters in England, Scotland, and Wales must provide their national insurance number, date of birth, and address when registering to vote~\cite{OSCEUK2015finalreport}.}. The information in the registration form is verified by checking the database of the Department of Works and Pensions (DWP), which makes the database a trusted part of the voting system~\cite{OSCEUK2015finalreport}. Polling station officials check eligibility by asking voter's name and address and comparing it with the information on the voter list~\cite{pollingstationhandbookuk2017}. Since 2023, voters in UK are required to show a photographic ID when voting in person\footnote{A list of valid photo ID-s is provided in \url{https://www.gov.uk/how-to-vote/photo-id-youll-need}}. 

\textbf{Verification of delivery}
The election observers, polling agents, and polling station officials are trusted to monitor how ballots are submitted into the ballot box~\cite{OSCEUK2005finalreport,OSCEUK2010finalreport}. 

\textbf{Verification of ballot box integrity}
The election observers, polling agents and polling station officials are trusted to monitor the ballot box\footnote{The procedures that the polling station staff must follow are described in~\cite{pollingstationhandbookuk2017}.}. Once the voting period ends, a polling station official fills in a ballot paper account, which can be used in the counting phase to check that the number of ballots in the ballot box matches the number of issued ballots~\cite{pollingstationhandbookuk2017}. The ballot box, the tendered ballots, and the list of serial numbers are sealed by the presiding clerk and optionally also by polling agents assigned by candidates or representatives of the candidates~\cite{lawcommissionmannerofvotinguk}. The ballot box is opened for counting, and once it is done, it is re-sealed. The sealed ballots, the tendered votes, and the list of serial numbers are forwarded to the registration officer of the area, who is trusted to store them for one year\footnote{The legislation does not specify the conditions for the retention of these items~\cite{lawcommissionmannerofvotinguk}.}. After one year, the stored election materials are destroyed~\cite{factsheetballotsecrecyuk2006}.

\textbf{Verification of tally integrity}
\label{paper_uk_tally_integrity}
Election observers, polling agents and polling station officials are trusted to observe the counting and report misconduct~\cite{OSCEUK2010finalreport}.

\subsection{Paper voting with cryptographic receipts}
Paper-based voting systems can be made cryptographically verifiable by printing cryptographic information on the ballots. The information can be presented in the form of candidate codes and ballot serial numbers, which are later posted to a public bulletin board for verification. However, care has to be taken to prevent the verification information on the ballots from being linked to voters, as this would violate ballot secrecy. 

Pr{\^e}t {\`a} Voter is a polling station-based voting system that provides end-to-end verifiability~\cite{PretAVoter3}. The ballot is two-sided, where the left side consists of a randomised candidate list, and the right side contains a ballot ID and cells where the voter can mark the preferences.

In this work, we refer to Culnane \emph{et al.}~\cite{pretavoteraustralia} version of Pr{\^e}t {\`a} Voter, which was used in the elections in the Australian state of Victoria. {Pr{\^e}t {\`a} Voter} differs from regular paper ballots used in Australian elections by its two-sided ballot printed by a print-on-demand printer in a private booth. The trust assumptions for achieving universal suffrage, equal suffrage, and verification of eligibility are the same as with Australian paper voting and are thus not covered in detail\footnote{While proof of identity is required for enrolling, there is no voter ID requirement for voting~\cite{australianvoteridreport2014}. The voter's eligibility is checked by polling station officials who ask voter's name, address, and whether the voter has already voted in this election.}.

\textbf{Ballot secrecy} 
It is assumed that the human-readable part of a ballot, which contains a candidate list and unique serial number corresponding to the voter’s receipt, is destroyed after voting\footnote{Using the human-readable part, an adversary could discover how the voter voted.}. Polling station officials are trusted to perform or/and enforce the shredding procedure after the voter has voted. In addition, the Electronic Ballot Marker (EBM), a computer assisting voters in ballot filling, is trusted not to leak the vote via side channels. It is assumed that the printer that retrieves and prints the appropriate ballot is resilient to side-channel attacks and kleptographic attacks. Also, votes on the Web Bulletin Board (WBB) stay private as long as there is at least one honest mixer and a threshold number of trustees sharing the decryption key do not collude. 

\textbf{Coercion-resistance} It is assumed that the vote is cast in a private setting where the coercer can not observe the voter. The voter is trusted not to record the voting process as the ballot contains cryptographic information along with the candidate list. However, Pr{\^e}t {\`a} Voter does not defend against pattern-based coercion attacks (a.k.a. Italian attack~\cite{Italianattacks}) or forced randomisation attacks.

\textbf{Verification of delivery}
Pr{\^e}t {\`a} Voter provides evidence of cast-as-intended verification and counted-as-cast verification if the voter performs a number of checks on the printed ballot and verifies the presence of their vote on the public part of WBB. Robustness and reliability of the private part of WBB depend on the honesty of a threshold of trustees handling it. The Australian implementation of Pr{\^e}t {\`a} Voter requires a threshold greater than two-thirds of trustees~\cite{pretavoteraustralia}.

\textbf{Verification of ballot box integrity}
Voters ought to perform a number of checks on the printed ballot and verify that their vote (code) is present on the public WBB. End-to-end verifiability relies on the assumption that sufficiently many voters run those checks. The means for guaranteeing the integrity of the physical ballot box are not part of Pr{\^e}t {\`a} Voter, and are assumed to be assured by honest polling station officials who may be monitored by election observers.

\textbf{Verification of tally integrity}
Every voter can verify the presence of their encrypted ballot on the public bulletin board. Moreover, anyone can verify that the published list of encrypted ballots corresponds to the announced results by downloading and checking a public electronic trace. It is assumed that a sufficient number of voters properly run those checks.

\section{Postal voting}
A common approach to providing voters with an option to vote remotely is to enable postal voting\footnote{\url{https://www.idea.int/news-media/news/special-voting-arrangements-svas-europe-country-postal-early-mobile-and-proxy}}. The importance of postal voting has been steadily increasing~\cite{whovotebypost2021}, but the Covid-19 pandemic accelerated the process as election organisers (EOs) had to introduce or expand remote voting options\footnote{\url{https://www.idea.int/news-media/news/elections-and-covid-19-how-special-voting-arrangements-were-expanded-2020}}. 

\subsection{Regular postal voting} 
While the introduction of postal voting may increase the participation rate, it can also affect the security guarantees provided by the election system. For example, it is not easy to ensure that every voter receives a postal ballot in a timely manner. As postal ballots can be filled in an uncontrolled environment, the possibility of voters being coerced can not be eliminated. Also, unless voters deliver the filled-in ballots to the polling station, they can not verify whether the ballots were correctly handled and reached the polling stations in time. 

For the sake of concreteness, we will again consider the postal voting system of the United Kingdom  in this paper.

The UK introduced on-demand postal voting in 2000~\cite{OSCEUK2005finalreport}. Since then, the ratio of postal voters has steadily increased. In the 2005 General Elections, 12\% of the votes were cast by mail. In the 2009 European Parliament elections, the percentage rose to 14\%, and in the 2019 General Elections, 21\% of votes were cast by mail~\cite{OSCEUK2010finalreport,ukelectionstatistics19182021}.

\textbf{Ballot secrecy}
Filled-in ballot is sealed into a ballot envelope, which is then placed inside a return envelope along with documents that are used to check voter's eligibility~\cite{OSCEUK2010finalreport}. Thus, the postal service and election organisers are trusted not use the information in the envelopes to violate ballot secrecy. To preserve ballot secrecy, postal ballots are mixed with regular ballots before counting~\cite{OSCEUK2010finalreport}. Election observers are trusted to check that the procedures are followed.

\textbf{Coercion-resistance} 
If the voters anticipate being coerced, they can use the option to cancel the request to vote by mail\footnote{To cancel the request to vote by mail, the voter must contact the local council at least 11 days before the elections. \url{https://www.electoralcommission.org.uk/i-am-a/voter/apply-vote-post\#paragraph-19876-title}}. Postal ballots are filled in an uncontrolled environment, which means that voters must be trusted to resist coercion and not to record the voting process. There are no special measures to mitigate the threat of coercion. 

\textbf{Universal suffrage}
The trust assumptions regarding voter registration are the same as for regular paper-based voting, see Section~\ref{paper_uk}. To apply for postal voting and when filling in the postal ballot, the voter must submit personal identifiers (signature, date of birth)~\cite{OSCEUK2010needsassessment}. The election organisers and polling station officials have to be trusted to validate the information provided by the voters.

\textbf{Equal suffrage}
The voter list contains marks for voters who have applied to use postal voting and thus can not vote in person~\cite{OSCEUK2010finalreport}. The polling station officials are trusted to keep the voter list up to date. They also have to be trusted to prevent postal voters from being allowed to cast a second vote.

\textbf{Verification of eligibility}
A postal vote must be accompanied by a filled-in postal voting statement that contains the voter's name, date of birth and signature\footnote{\url{https://www.electoralcommission.org.uk/sites/default/files/pdf_file/Making-Your-Mark-Example-Postal-Voting-Statement-GB-English-A4.pdf}}. Starting from 2015, it has been mandatory for returning officers to verify the personal identifiers on all returned postal ballots~\cite{OSCEUK2015finalreport}. The verification is usually done with the help of specific software, which has to be trusted\footnote{The verification system can not detect fake identifiers if they are used both on the postal voting application and on the statement accompanying the filled-in ballot~\cite{OSCEUK2010finalreport}.}. If the software does not find a match, a human operator has to review the identifiers, which means that the human operator has to be trusted.

\textbf{Verification of delivery}
Voters have to trust that the ballots are delivered on time\footnote{Tracking of postal votes was trialled in 2006 in some parts of the UK~\cite{postalvotetrackingtrialuk2006}.}. If a voter does not trust the postal service, there is also a possibility to return the ballot to the polling station~\cite{OSCEUK2010finalreport}. Postal ballots can also be collected and delivered by third parties, including political parties, which must be trusted to behave honestly~\cite{OSCEUK2010finalreport}. However, there is a plan to introduce new legislation in the UK restricting political campaigners from handling postal ballots~\cite{ukelectionbill2022}.

\textbf{Verification of ballot box integrity}
Postal ballots must be securely stored. The election officials responsible for handling the postal ballots must be trusted not to violate ballot secrecy and to assure the integrity of the ballots. Accredited election observers can monitor how the postal ballots are received~\cite{OSCEUK2017needsassessment}. However, it is unclear how the stored ballots are secured and whether election observers can also monitor that the ballots are properly stored. If ballots are stored in a storage facility, the operator of that facility would also be part of the trust base.

\textbf{Verification of tally integrity}
The means for assuring tally integrity are the same as those used for regular paper-based ballots, see Section~\ref{paper_uk_tally_integrity}.

\subsection{Postal voting with cryptographic receipts}
STROBE~\cite{STROBE} is a framework that helps to achieve end-to-end verifiability in postal voting systems. The framework was introduced by Josh Benaloh in 2021 and has not yet been implemented in practice. 

STROBE differs from regular postal voting as each voter receives two ballots, one for voting and the other for verification\footnote{We cover the version of STROBE, where voters receive two ballots.}. These ballots contain cryptographic information in the form of a ballot ID and short selection codes for each candidate\footnote{The selection codes are generated by the election organiser from pre-computed encryptions of each candidate on a ballot. The ballot ID is formed by hashing all the encryptions for that ballot. The election organiser publishes encryptions, short selection codes, and ballot ID for each ballot to the public bulletin board.}. The idea is to make it unpredictable which ballot is verified. Thereby, there is a high likelihood of detecting tampered ballots during verification, giving indirect guarantees about the integrity of the election result.

As STROBE is not a voting system but a framework to add verifiability to a voting system, the assumptions for achieving universal suffrage, equal suffrage, and verification of eligibility are not treated separately in this section.

\textbf{Ballot secrecy} The printing house is trusted not to retain information about ballots. The postal service and election organiser are trusted not to violate ballot secrecy by illegally storing ballot ID, the connection between selection codes and candidates, and linking with the voter's identity. Such information could be used to query the public bulletin board to learn voter's choices. Additionally, it is assumed that the threshold of decryption key holders will not collude. 

\textbf{Coercion-resistance} STROBE is not designed to mitigate coercion. Therefore, voters must be trusted to evade coercion and not record the voting process.  

\textbf{Verification of delivery} After receiving the ballots, the election organiser publishes selection codes corresponding to the voters' choices on the public bulletin board. Every voter can verify whether their chosen selection codes appear on the bulletin board, and it is assumed that a sufficient number of voters do this. The election organiser must be trusted to correctly publish the  codes.

\textbf{Verification of ballot box integrity} The election organiser publishes ballot ID-s and chosen selection codes for all the received ballots. It is assumed that a sufficient number of voters will perform verification to detect a systematic removal or modification of ballots. Election observers or auditors are trusted to check that the number of voters matches the number of ballots in the ballot box and the number of published receipts.

\textbf{Verification of tally integrity} After the voting phase has ended, encryptions of cast votes are homomorphically combined, and the tally is verifiably decrypted. Anyone can verify correctness of the final tally by aggregating the encryptions published on the bulletin board and verifying the proof of decryption correctness. It is assumed that at least some voters can and are willing to do such checks. It is also assumed that a sufficient number of voters verify their votes, hence showing that the information on the public bulletin board is valid.

\section{Voting via voting machines}
Voting machines are devices which collect and tally votes. Some of these machines let the voters directly enter their choices, while others scan filled-in ballots. If the source code of the voting machine is not available and the machine does not produce a paper trail, the machine can be viewed as a black box that has to be trusted to function correctly. Some voting machines produce a paper trail, making it possible to conduct post-election risk-limiting audits~\cite{DBLP:journals/ieeesp/LindemanS12}.
    
\subsection{Voting machines used in Bulgaria}
In the following, we are describing the deployment of voting machines in Bulgaria\footnote{Smartmatic A4-517 machines were used: \url{https://www.smartmatic.com/media/article/smartmatic-supports-24-elections-in-11-countries-in-2021/}.}. The information is based on OSCE reports and an interview with an election observer~\cite{OSCEBulgaria2021julyfinalreport,OSCEBulgaria2021aprilfinalreport,Bulgaria2021Jan}.

The first country-wide trial of using voting machines was conducted during the parliamentary elections in April 2021. At that time, the voters could choose to cast a paper ballot or vote via a voting machine. Three months later, during the early parliamentary elections, it was mandatory to use voting machines in polling stations with over 300 registered voters.

\textbf{Ballot secrecy}
The voting machine does not learn the voter's identity\footnote{The voter uses a smart card provided by the polling station official to cast a vote.}. The vendor is trusted to provide voting machines that fulfil the requirement, whereby it should be impossible to determine in which order the votes were stored on the external memory device. The vendor is trusted to use the same components in the voting machines as only a few voting machines were tested regarding electromagnetic emanations. Once a vote is cast, the voter's choice is printed on a receipt. The voter is assumed to fold it and put it into a ballot box. 

\textbf{Coercion-resistance}
Voters' privacy is protected with the help of small privacy screens installed on both sides of the voting machine. Thus, the voter has to trust that cameras have not been placed on the premises of the polling station and that the voting machines have been positioned in a manner preventing third parties from viewing the voting machine's screen. Voters are trusted not to record the voting process. However, the lack of a polling booth makes it more difficult for the voters to secretly record the voting process\footnote{Vote buying was one of the main reasons for introducing voting machines. Polling station officials were instructed to monitor that voters do not use recording devices.}.

\textbf{Universal suffrage}
All citizens with a permanent address in Bulgaria are automatically registered. Prior registration is not required to vote abroad. The National Population Register has to be trusted to provide valid data, as it is used to compile voter lists. To vote, the voters need to provide an identity document to a polling station official.

\textbf{Equal suffrage}
In general, voters can vote only in their designated polling station\footnote{There are no such restrictions when voting abroad, which could lead to double voting.}. However, in some rare cases, unregistered voters are allowed to register in the polling station on the election day, which may result in multiple registration and double voting. An eligible voter is provided with a smart card, allowing to use a specific voting machine. Polling station officials are trusted to check that each voter gets only one smart card, as two subsequent votes on the same machine must be cast with different smart cards. To prevent the same voter from later casting a second vote, the polling station official is trusted to keep the voter's identity document until the voter returns the smart card, submits the receipt to the ballot box, and adds a signature to the voter list.

\textbf{Verification of eligibility}
The polling station employee is trusted to check the voter's eligibility by requesting an identity document and comparing the information on the document with the information on a paper-based voter list.

\textbf{Verification of delivery}
Election organiser is trusted to have properly configured the voting machines. The voter is assumed to verify that the cast vote matches the vote printed on the receipt. The voter submits the receipt to the ballot box. The polling station officials must be trusted to properly seal the USB drives, receipts, and results printed by voting machines into a tamper-evident envelope and deliver the envelope to the district counting centre. 

\textbf{Verification of ballot box integrity}
As the machines are air-gapped, they are trusted to provide valid signing times when votes are signed. The integrity of the voting machine's software is verified before installation by checking the hash value of the installation file\footnote{Observers were not able to monitor the configuration process of the voting machines.}. It was possible to verify the integrity of the installed software by booting a special USB containing a hash checker. Support for the voting machines was provided by local contractors, who had to be trusted not to tamper with the machines\footnote{More than 11000 voting machines were used across 9398 polling stations.}. Polling station officials had to be trusted not to tamper with the voting machines and ballot boxes containing the receipts. In case observers are present, they are trusted to behave honestly.

\textbf{Verification of tally integrity}
Post-election audits were performed in some polling stations to check whether counting the receipts gives the same results as the tally provided by the voting machines. The election organiser is trusted not to be malicious when deciding which polling stations to audit, as the set of audited polling stations was not chosen randomly. Digitally signed votes were published.

\section{Internet voting}
Internet voting (i-voting) systems enable voting in an uncontrolled environment, which may impact both vote privacy and vote integrity. Therefore, they usually provide voters with means to avoid coercion while allowing voters to verify that their votes were correctly handled. However, an optimal balance has to be found as there is an inherent conflict between these security requirements. To provide coercion-resistance, it must not be possible to prove how the vote was cast, making it challenging to provide universal verifiability in the remote setting. 

In this section, we review the trust assumptions of the Estonian i-voting system IVXV, which provides individual verifiability and coercion-resistance.

\subsection{Estonian IVXV}
The updated version of the Estonian i-voting system (code-named IVXV) was introduced in 2017~\cite{DBLP:conf/voteid/HeibergMVW16,IVXV2017eng,ivxvarhitektuur}. While there were few changes to the client-side, the sever-side was completely rewritten. The collection of ballots was distributed between Vote Collectors and a Registration Service independent of the election organiser. Once the voting period is over, the ballots are processed and anonymised by a Ballot Box Processor. A mix-net was introduced to anonymise votes, decryption was enhanced by providing proofs of correct decryption, and the election specific private key was threshold secret-shared between the trustees.

\textbf{Ballot secrecy}
The voting device, verification device, verification application, and voting application are trusted not to violate ballot secrecy. Thus, the voter is trusted to ensure that the voting and verification devices are secure. The server-side processes encrypted ballots and the election organiser is trusted to anonymise the ballots before the tallying phase. Observers are assumed to check that ballots are anonymised. It is assumed that encryption can not be broken. Thus, to violate ballot secrecy, the election-specific decryption key would have to leak during key generation, or at least 5 of 9 keyholders would have to collude.

\textbf{Coercion-resistance}
The possibility to re-vote makes it impossible to prove whether the cast vote reaches the tallying phase\footnote{I-vote can also be invalidated by casting a paper vote in a polling station.}. For the mitigation measure to be effective, the voter must be aware and willing to use the option to re-vote. It is assumed that the coercer does not have control over the voter's voting device as otherwise it would be possible to detect whether the voter complies or casts a re-vote. It is also assumed that the election organiser does not collude with the coercer as the Vote Collector and Ballot Box Processor can see which voters re-voted. Similar information can be inferred by the party providing Vote Collector with OCSP responses regarding the validity of voters' signing certificates~\cite{IVXVprotocolseng}.

\textbf{Universal suffrage}
Eligible voters are automatically included in the voter list, which is compiled based on the  electronic population registry. The population registry has to be trusted to provide valid information.

\textbf{Equal suffrage}
The election observers or auditors are trusted to verify that only the last vote cast by each voter gets tallied. Third parties must not be able to use the electronic identity tokens (eIDs) belonging to other voters. Thus, it is assumed that malicious usage of voters' eIDs would be detected. 

\textbf{Verification of eligibility}
Eligibility is checked by requesting the voters authenticate to the voting system by using either an ID card or Mobile-ID. The state provided eIDs are trusted to be secure. After the elections, the Ballot Box Processor verifies that only eligible voters were able to vote. Election observers or auditors are assumed to check that the Ballot Box Processor functions properly.

\textbf{Verification of delivery}
The voter is assumed to use a smartphone-based vote verification application to fetch the cast ballot from the Vote Collector. The vote verification application reads the contents of the QR code displayed by the voting application to initiate vote verification. Thus, the voter has to trust that the voting application and verification application do not collude maliciously.

\textbf{Verification of ballot box integrity}
The trust assumptions used for cast-as-intended verification are expanded by having to trust that third parties cannot use voters' eID-s to re-vote as this would not be detected by the vote verification application~\cite{estoniannextsteps2020}. The Vote Collector and Registration Service are trusted not to collude, as otherwise it would be possible to drop ballots~\cite{DBLP:conf/voteid/HeibergMVW16}.

\textbf{Verification of tally integrity}
The Vote Collector and Registration Service are trusted not to collude. It is assumed that at least some election observers and auditors are not malicious. They have to verify that only eligible voters could vote, that the set of ballots collected by the Vote Collector matches the set of ballots registered by the Registration Service, and that the ballots were properly anonymised and decrypted. Some of these tasks rely on verification software. Thus, the software used to verify mixing proofs and decryption proofs can be viewed as trusted components of the voting system.

\section{Discussion and conclusions}
Our original motivation behind this research was to compare the security levels of different voting methods on a common scale. If method $A$ would need a strictly smaller set of trust assumptions than method $B$, then we could say that $A$ is in this sense more secure than $B$. 

During our research we concluded that, besides the mere observation that a trust assumption is needed, we also have to consider the impact in case a particular trust assumption is violated. Thus we introduced three levels of impact criticality. In the lightest case, only one vote is affected, and this probably does not change the final outcome of elections. In the mid-severe case, a subset of votes are affected, and this has a potential of changing the outcome of elections by a few seats in the representative body to be elected. And finally, in the most critical case, all the votes may be impacted as a result of the attack enabled due to the failing trust assumption.

\subsection{Overview of the trust assumptions}
In the following, we generalise the results of our case studies. The identified trust assumptions are summarised in Table~\ref{tab:ivoting_threats_corrupted_parties_new}. 

\textbf{Ballot secrecy}
Voters are assumed not to disclose the details of their voting process, for example, by not recording the process. In addition, it is assumed that the environment where the ballot is filled does not contain recording equipment. For example, in case of i-voting systems, the voting device has to be trusted unless code voting is used, and in case of machine voting, the hardware vendor is trusted to deliver machines that do not have side-channel leakages. Polling station officials are trusted not to abuse their position to violate ballot secrecy. For example, in the case of postal voting, it is assumed that they will not link voter's identity to the vote on the ballot. Similarly, postal voting systems rely on the trustworthiness of the deliverer as they could violate ballot secrecy by opening the double envelopes. The voting schemes that use printed cryptographic information are dependent on the software generating the information and on the printing facility, as the cryptographic information must not leak and must not be linked to the voters who receive the information. When considering voting schemes that rely on cryptography, it is common for the election organiser to distribute the trust regarding key management and anonymisation of ballots.

\textbf{Coercion-resistance}
In general, trust assumptions regarding ballot secrecy can be used as a basis for deriving some of the trust assumptions for coercion-resistance. If ballot secrecy can be violated, it is possible to coerce the voters. While the polling booth used to be instrumental for providing coercion-resistance, the proliferation of small recording devices has made it practically impossible to prevent the voting process from being recorded~\cite{DBLP:conf/uss/Benaloh13}. While i-voting has the potential to provide the highest level of coercion-resistance by giving the voter the option to mitigate coercion by casting a re-vote or by voting with fake credentials, the coercion-resistance of i-voting schemes can rely on multiple trust assumptions. For example, the voter would have to be aware of the way how the coercer is monitoring the voter to apply the countermeasures. The voter may not be able to detect that the computer is compromised and used to monitor how the vote is cast. Also, even if the anti-coercion measures (like the option of re-voting) are implemented by the system, the voter has to be aware of them and should be willing to use them. It becomes very difficult to provide coercion-resistance when forced abstention attacks are included in the threat model, as multiple parties may learn, which voters participated in the elections. These may include the election organiser, polling station officials, and identity provider.

\textbf{Equal and universal suffrage}
While integrity of the electoral register has to be trusted to guarantee universal suffrage, voter registration and universal suffrage are not specific to a voting system, but depend on the  legislation. Still, the voter must be given the possibility to vote. For example, voting machine vendors must provide reliable hardware and software that is usable by the voters. When considering equal suffrage, double voting is commonly prevented by relying on the procedures and proper marking of the voter list by polling station officials. Thus, polling station officials and the system for managing the voters list must be trusted. Also, election observers and auditors must be trusted to check that the procedures are followed. However, election observers can not check whether the voting machine's software correctly represents voters' preferences. While risk limiting audits can be used after the elections to check that the voting machines did not tamper with the results, this option is only available for voting machines that provide a paper trail. For voting machines that do not provide a paper trail, the software vendor has to be trusted. In the case of i-voting, it has to be assumed that third parties can not vote with someone else's credentials without being detected, which means that voters and identity providers have to be trusted to properly handle the voting credentials. In addition, the software vendor of an i-voting system may have to be trusted, unless the voting system is software independent. In the case of postal voting, the postal service must be trusted not to drop votes of their choice. 

\textbf{Verification of eligibility}
Verification of eligibility relies on the assumption that voters provided valid information when registering to vote. As the registrar is responsible for verifying voters' identities and including them in the voter list, the registrar is a trusted part of the election system. For polling station-based voting systems, voter eligibility is checked by polling station officials. In the case of postal voting systems, eligibility can be checked by comparing the information provided by the voter during the registration with the information provided in the ballot. This can be done manually by the polling station officials or the process can be automated with software. If the software is used for verifying eligibility, it must be considered as a part of the trust base. The way how i-voting systems identify voters varies significantly, but the election organiser is trusted to ensure that ballots are accepted only from eligible voters. In addition, the election observers are also responsible for verifying that the i-voting system properly checks eligibility. 

\textbf{Verification of delivery}
When the vote is cast using postal voting, the postal service has to be trusted not to drop the votes. If the postal ballots are delivered to the polling station, the polling station employees along with the election observers are responsible for correctly handling the received ballots. In the case of regular paper voting, there are no trust assumptions for delivering the ballot as it is the voter who submits the ballot to the ballot box. However, with cryptographic paper voting, the printing house may misprint the cryptographic information, which can invalidate the ballot or prevent it from being verified. The same holds for cryptographic postal voting. In general, when the voter is provided with the means to verify whether the ballot was correctly encoded or delivered, the voter is trusted to perform this check. In the case of i-voting, a voter's computer has to be trusted not to prevent the vote from being verified. In addition, the software vendor of the i-voting system may have to be trusted, unless the voting system is software independent. The trust assumptions for machine voting differ depending on whether the machine provides a paper trail. If a paper trail is provided, the voter is responsible for checking the receipt. However, if no paper trail is provided, the software running on the voting machine must be trusted to correctly store voters' preferences.

\textbf{Verification of ballot box integrity}
Election observers and auditors are trusted to monitor or verify the integrity of the ballot box. Voting machines that do not produce a paper trail are an exception as integrity of their ballot boxes can not be easily monitored nor verified. Thus, polling station officials and infrastructure providers have to be trusted not to tamper with such voting machines. For the voting machines that provide a paper trail, the infrastructure provider has to be trusted not to damage or tamper with the paper receipts. In this case, also election auditors are trusted to properly conduct post election risk limiting audits. When considering i-voting systems that only provide individual verifiability, the election organiser has to be trusted along with the infrastructure provider, election auditor, and software vendor as the general public can not verify how the ballot box is processed. However, for i-voting systems that provide universal verifiability, in principle, anyone could verify the integrity of the ballot box, but at minimum, the verification is expected to be conducted by experts in the role of election auditors. In the case of regular paper ballots and postal ballots, polling station officials and storage facility employees have to be trusted. 

\textbf{Verification of tally integrity}
In the case of regular paper-based voting systems and regular postal voting systems, polling station officials have to be trusted to count the ballots correctly. In the case of machine voting, the voting machines can export digital ballot boxes. When no paper trail is provided, the software running on the voting machine has to be trusted to export a ballot box that represents the preferences of the voters who cast the votes. The digital ballot boxes may be delivered to a central location for tallying. In such cases, the deliverer must be trusted along with the polling station officials and election organisers. If the machines provide a paper trail, the election organiser must be trusted to use proper statistical means to conduct risk-limiting audits to ensure that even if some voting machines behaved incorrectly, it did not affect the election result. 
For i-voting systems that only provide individual verifiability, the election organiser has to be trusted along with the infrastructure provider and software vendor as the general public can not verify how tallying is done. Therefore, election observers and auditors play a significant role in such voting systems and have to be trusted. For i-voting systems that provide universal verifiability, in principle, anyone could verify the tally, but at minimum the verification is expected to be conducted by experts in the role of election auditors. Regardless of the voting system, election observers are trusted to monitor how the votes are counted. If auditors rely on software to verify cryptographic proofs, these software components must be considered to be part of the trust base.

\subsection{Comparative analysis of the trust assumptions}

In this section we present the conclusions of our analysis in terms of the amount and criticality of the studied trust assumptions.

\textbf{Ballot secrecy}
Paper-based voting in a polling station relies on the least amount of trust assumptions when considering ballot secrecy. As it is a well-established method of voting, many conceivable attacks against ballot secrecy have been attempted in the past and hence have counter-measures in place. 

Cryptographic paper voting and cryptographic postal voting have the largest number of trust assumptions and three out of them are critical. All the ballots must be kept confidential between their creation and vote casting; secrecy of the vote can then be satisfied only if there exists a chain of trusted entities throughout the vote casting process. 

Similarly, i-voting requires the same number of critical trust assumptions due to the necessity to trust the election organiser and software vendor with the security implementation of the voting process. Furthermore, we need to trust devices used for casting and verifying the votes.

\textbf{Coercion-resistance}
Among the voting methods, paper-based voting and voting via voting machines rely on the fewest security assumptions regarding coercion-resistance. As stated above, paper-based voting relies on the least number of trust assumptions when it comes to ballot secrecy. Coercers have only a few ways of knowing whether their efforts have succeeded in making voters cast their votes in a specific manner. 

Cryptographic postal voting requires the largest number of trust assumptions, three of which may affect all the votes, and four of which may affect a subset of the votes. As a result of several parties having the possibility to violate ballot secrecy, they might influence the way the voters express their preferences.

I-voting has the largest number of critical trust assumptions. Even though some i-voting schemes provide means to counter coercion, they cannot protect from all possible attacks (like family voting). This shows that i-voting is a more suitable method for communities where coercive practices are uncommon.

\textbf{Equal suffrage}
As both regular and cryptographic paper voting rely on existing identification mechanisms to check for voter eligibility, they require the fewest trust assumptions.

I-voting has the highest number of trust assumptions, four of which are critical. Those assumptions are needed as any i-voting system which is used in large-scale elections requires robust infrastructure, an efficient and reliable identification system, an online population registry, and an honest election organiser. 

\textbf{Verification of eligibility}
The results of our analysis suggest that i-voting provides the weakest guarantees regarding voters' eligibility since there are four critical trust assumptions. In contrast, voting in a polling station has the potential to provide the highest level of guarantees by requiring the checking of the voters' identity documents. 

\textbf{Verification of delivery}
Paper-based voting does not entail any trust assumption in terms of verifying that the ballots have been delivered. Voters do not need to trust anyone since they deliver their ballot to the ballot box themselves. 

Methods that provide individual verifiability (cryptographic paper voting, cryptographic postal voting, i-voting with individual verifiability) require more trust assumptions. Firstly, voters are trusted to perform verification, which can sometimes be confusing. Furthermore, printing houses should be trusted for cryptographic postal and paper voting, since incorrect codes can prevent voters from verifying their votes. In the case of i-voting, a voter's computer and software vendor should be trusted not to prevent the vote from being verified.

\textbf{Verification of ballot box integrity}
Methods that provide universal verifiability have the fewest trust assumptions related to the verification of ballot box integrity. Among these methods are cryptographic paper voting, cryptographic postal voting, and i-voting with universal verifiability. I-voting systems with individual verifiability have the greatest number of trust assumptions, all of which are critical. Individual verifiability does not guarantee the integrity of all other ballots submitted by all voters, but only provides assurance of the integrity of the ballot submitted by a particular voter. 

\textbf{Verification of tally integrity}
Methods that provide universal verifiability require the least amount of trust assumptions for verification of tally integrity. The reason for this is that several entities have access to special cryptographic techniques that can be used to verify the final tally.

On the other hand, machine voting with no paper trail requires the highest number of assumptions with two assumptions being critical. As a result, there is need to trust software, election organisers, infrastructure provider, and delivery services to accomplish verification of tally integrity.

\subsubsection{Acknowledgements}  This paper has been supported by the Estonian Research Council under the grant number PRG920.

\bibliographystyle{splncs04}
\bibliography{references}

\begin{table*}[ht]

\centering
\caption{An overview of trust assumptions for different types of voting systems.}
\label{tab:ivoting_threats_corrupted_parties_new}
\begin{minipage}{\textwidth}
\resizebox{\textwidth}{!}{%

\begin{tabular}{|l|l|P{0.65cm}|P{0.65cm}|P{0.65cm}|P{0.65cm}|P{0.65cm}|P{0.65cm}|P{0.65cm}|P{0.65cm}|P{0.65cm}|P{0.65cm}|P{0.65cm}|P{0.65cm}|}
\hline
 &&
  \rotatebox{90}{\textbf{Voter}} &
  \rotatebox{90}{\textbf{Voter's computer}} &
  \rotatebox{90}{\textbf{Registrar}} &
  \rotatebox{90}{\begin{tabular}{p{3.5 cm}}\textbf{Election organiser \newline or election services}\end{tabular}} &
  \rotatebox{90}{\begin{tabular}{p{3.5 cm}}\textbf{Infrastructure} \textbf{provider$\backslash$maintenance }\hspace{1mm}\end{tabular}} &
  \rotatebox{90}{\begin{tabular}{l}\textbf{Polling station} \textbf{official}\end{tabular}} &
  \rotatebox{90}{\begin{tabular}{l}\textbf{Printing house}\end{tabular}} &
  \rotatebox{90}{\begin{tabular}{p{3.5 cm}}\textbf{Election} \textbf{observer$\backslash$auditor}\end{tabular}} &
  \rotatebox{90}{\begin{tabular}{l}\textbf{Identity provider}\end{tabular}} &
  \rotatebox{90}{\begin{tabular}{p{3.5 cm}}\textbf{Postal service or} \textbf{third party delivery}\hspace{1mm} \end{tabular}} &
  \rotatebox{90}{\textbf{Hardware vendor}} &
  \rotatebox{90}{\begin{tabular}{l}\textbf{Software vendor }\end{tabular}} \tn \hlineB{2}

\multirow{9}{*}{\begin{tabular}[c]{@{}l@{}}\textbf{Ballot}\tn \textbf{secrecy}\end{tabular}}
    & Paper voting      &\CIRCLE & &  &  &
    \darkgraycell{footnote}\CIRCLE\tablefootnote[7]{} %\diagfil{0.68cm}{gray!25}{white}{\hspace{0.65em}\CIRCLE$^7$} 
    &  \darkgraycell{}\CIRCLE
    %\diagfil{0.68cm}{gray!25}{white}{\CIRCLE} 
    &  
    & \darkgraycell{footnote}\CIRCLE\tablefootnote[9]{}
    %\diagfil{0.68cm}{gray!25}{white}{\hspace{0.40em}\CIRCLE$^\text{9}$} 
    &  &  &  &    \tn \hhline{|~|*{13}{-}} %\cline{2-14}
    & Cryptographic paper voting &\CIRCLE  & &  & \graycell{footnote}\CIRCLE\tablefootnote[1]{} &
    \darkgraycell{footnote}\CIRCLE\tablefootnote[7]{}
    %\diagfil{0.68cm}{gray!25}{white}{\hspace{0.65em}\CIRCLE$^7$} 
    & \darkgraycell{}\CIRCLE
    %\diagfil{0.68cm}{gray!25}{white}{\CIRCLE}  
    & \graycell{footnote}\LEFTcircle\tablefootnote[11]{} 
    & \darkgraycell{footnote}\CIRCLE\tablefootnote[9]{}
    %\diagfil{0.68cm}{gray!25}{white}{\hspace{0.40em}\CIRCLE$^\text{9}$} 
    &  &  &  &    \tn \hhline{|~|*{13}{-}} %\cline{2-14}
    & Postal voting     &\CIRCLE &  &  &  &
    \darkgraycell{footnote}\CIRCLE\tablefootnote[4]{}
    %\diagfil{0.68cm}{gray!25}{white}{\hspace{0.65em}\CIRCLE$^4$}
    & \darkgraycell{}\CIRCLE
    %\diagfil{0.68cm}{gray!25}{white}{\CIRCLE}  
    &  & \darkgraycell{footnote}\CIRCLE\tablefootnote[9]{}
    %\diagfil{0.68cm}{gray!25}{white}{\hspace{0.40em}\CIRCLE$^{\text{9}}$} 
    &  & \darkgraycell{}\CIRCLE
    %\diagfil{0.68cm}{gray!25}{white}{\CIRCLE}  
    &  &     \tn \hhline{|~|*{13}{-}} %\cline{2-14}
    & Cryptographic postal voting  &\CIRCLE &  &  &\graycell{footnote}\CIRCLE\tablefootnote[1]{}  &
    \darkgraycell{footnote}\CIRCLE\tablefootnote[4]{}
    % \diagfil{0.68cm}{gray!25}{white}{\hspace{0.65em}\CIRCLE$^4$}
    & \darkgraycell{}\CIRCLE
    %\diagfil{0.68cm}{gray!25}{white}{\CIRCLE}  
    &\graycell{}\CIRCLE &
    \darkgraycell{footnote}\CIRCLE\tablefootnote[9]{}
    %\diagfil{0.68cm}{gray!25}{white}{\hspace{0.40em}\CIRCLE$^{\text{9}}$} 
    &  & \darkgraycell{}\CIRCLE
    %\diagfil{0.68cm}{gray!25}{white}{\CIRCLE}  
    &  &  \graycell{}\LEFTcircle   \tn \hhline{|~|*{13}{-}} %\cline{2-14}
    & Machine voting  &\CIRCLE &  &  &  & \darkgraycell{footnote}\CIRCLE\tablefootnote[7]{} %\diagfil{0.68cm}{gray!25}{white}{\hspace{0.65em}\CIRCLE$^7$} 
    & \darkgraycell{}\CIRCLE
    %\diagfil{0.68cm}{gray!25}{white}{\CIRCLE}  
    &  & \darkgraycell{footnote}\CIRCLE\tablefootnote[9]{}
    %\diagfil{0.68cm}{gray!25}{white}{\hspace{0.40em}\CIRCLE$^\text{9}$} 
    &  &  & \graycell{}\LEFTcircle &   \tn \hhline{|~|*{13}{-}} %\cline{2-14}
%    & I-voting with code voting   &\CIRCLE  & &  &\CIRCLE  &  &  &\CIRCLE  &\CIRCLE  &  &\CIRCLE  &  &\CIRCLE   \tn  \cline{2-14}
    & I-voting   &\CIRCLE &$\;\,$\CIRCLE\tablefootnote[5]{}  &  &\graycell{footnote}\CIRCLE\tablefootnote[1]{}  &  &  &  & \hspace{0.4em}\graycell{}\CIRCLE\tablefootnote[9]{}  &  &  &  &\graycell{footnote}\CIRCLE\tablefootnote[5]{}   \tn
    \hlineB{2}
    
\multirow{9}{*}{\begin{tabular}[c]{@{}l@{}}\textbf{Coercion-}\tn \textbf{resistance}\end{tabular}}  
    & Paper voting      &\CIRCLE &  &  &  & \darkgraycell{footnote}\CIRCLE\tablefootnote[7]{} %\diagfil{0.68cm}{gray!25}{white}{\CIRCLE$^7$} 
    & \darkgraycell{footnote}\CIRCLE\tablefootnote[10]{}
    %\diagfil{0.68cm}{gray!25}{white}{\hspace{0.71em}\CIRCLE$^{\text{10}}$}   
    &  &  \darkgraycell{footnote}\CIRCLE\tablefootnote[9]{}
    %\diagfil{0.68cm}{gray!25}{white}{\hspace{0.4em}\CIRCLE$^{\text{9}}$}  
    &  &  &  &  \tn  \hhline{|~|*{13}{-}} %\cline{2-14}
    & Cryptographic paper voting      &\CIRCLE &  &  & \graycell{}\CIRCLE  & \darkgraycell{footnote}\CIRCLE\tablefootnote[7]{} %\diagfil{0.68cm}{gray!25}{white}{\CIRCLE$^7$}  
    & \darkgraycell{footnote}\CIRCLE\tablefootnote[10]{}
    %\diagfil{0.68cm}{gray!25}{white}{\hspace{0.71em}\CIRCLE$^{\text{10}}$} 
    & \graycell{footnote}\LEFTcircle\tablefootnote[11] & \darkgraycell{footnote}\CIRCLE\tablefootnote[9]{} %\diagfil{0.68cm}{gray!25}{white}{\hspace{0.4em}\CIRCLE$^{\text{9}}$}  
    &  &  &  & \graycell{footnote}\LEFTcircle\tablefootnote[11]{}  \tn  \hhline{|~|*{13}{-}} %\cline{2-14}
    & Postal voting     &\CIRCLE &  &  &  & \darkgraycell{footnote}\CIRCLE\tablefootnote[4]{}
    %\diagfil{0.68cm}{gray!25}{white}{\CIRCLE$^4$}  
    & \darkgraycell{}\CIRCLE
    %\diagfil{0.68cm}{gray!25}{white}{\CIRCLE}  
    &   & \darkgraycell{footnote}\CIRCLE\tablefootnote[9]{}
    %\diagfil{0.68cm}{gray!25}{white}{\hspace{0.4em}\CIRCLE$^{\text{9}}$} 
    &  & \darkgraycell{}\CIRCLE
    %\diagfil{0.68cm}{gray!25}{white}{\CIRCLE} 
    & &   \tn \hhline{|~|*{13}{-}} %\cline{2-14}
    & Cryptographic postal voting     &\CIRCLE &  &  & \graycell{}\CIRCLE  & \darkgraycell{footnote}\CIRCLE\tablefootnote[4]{}
    %\diagfil{0.68cm}{gray!25}{white}{\CIRCLE$^4$}  
    & \darkgraycell{}\CIRCLE
    %\diagfil{0.68cm}{gray!25}{white}{\CIRCLE} 
    &\graycell{}\CIRCLE  & \darkgraycell{footnote}\CIRCLE\tablefootnote[9]{} %\diagfil{0.68cm}{gray!25}{white}{\hspace{0.4em}\CIRCLE$^{\text{9}}$}  
    &  & \darkgraycell{}\CIRCLE %\diagfil{0.68cm}{gray!25}{white}{\CIRCLE}  
    &  & \graycell{}\LEFTcircle  \tn  \hhline{|~|*{13}{-}}
    & Machine voting  &\CIRCLE &  &  &  & \darkgraycell{footnote}\CIRCLE\tablefootnote[7]{} %\diagfil{0.68cm}{gray!25}{white}{\CIRCLE$^7$} 
    & \darkgraycell{footnote}\CIRCLE\tablefootnote[10]{}
    %\diagfil{0.68cm}{gray!25}{white}{\hspace{0.71em}\CIRCLE$^{\text{10}}$}  
    &  & \darkgraycell{footnote}\CIRCLE\tablefootnote[9]{}
    %\diagfil{0.68cm}{gray!25}{white}{\hspace{0.4em}\CIRCLE$^{\text{9}}$}  
    &  &  &  &    \tn \hhline{|~|*{13}{-}}
%    & I-voting with code voting   &\CIRCLE &  &  &\CIRCLE  &  &  &\CIRCLE  &  &  &  &  &   \tn  \cline{2-14}
    & I-voting &\CIRCLE &$\;\,$\CIRCLE\tablefootnote[5]  &  &\graycell{}\CIRCLE  &  &  &  & \hspace{0.40em}\graycell{}\CIRCLE\tablefootnote[9]{} &\graycell{footnote}\CIRCLE\tablefootnote[10]{}  &  &  &\graycell{footnote}\CIRCLE\tablefootnote[5]{}   \tn  
    \hlineB{2}
     
\multirow{9}{*}{\begin{tabular}[c]{@{}l@{}}\textbf{Equal $\&$}\tn
\textbf{universal}\tn
\textbf{suffrage}\end{tabular}}  
    & Paper voting      & &  &\graycell{}\CIRCLE  &  &  & \darkgraycell{}\CIRCLE
    %\diagfil{0.68cm}{gray!25}{white}{\CIRCLE}  
    &  & \darkgraycell{footnote}\CIRCLE\tablefootnote[9]{}
    %\diagfil{0.68cm}{gray!25}{white}{\hspace{0.4em}\CIRCLE$^{\text{9}}$}  
    &  &  &  &   \tn  \hhline{|~|*{13}{-}} %\cline{2-14}
    & Cryptographic paper voting      & &  &\graycell{}\CIRCLE  &  &  & \darkgraycell{}\CIRCLE
    %\diagfil{0.68cm}{gray!25}{white}{\CIRCLE} 
    &  & \darkgraycell{footnote}\CIRCLE\tablefootnote[9]{}
    %\diagfil{0.68cm}{gray!25}{white}{\hspace{0.4em}\CIRCLE$^{\text{9}}$}  
    &  &  &  &   \tn  \hhline{|~|*{13}{-}} %\cline{2-14}
    & Postal voting     & &  &\graycell{}\CIRCLE  &  &  & \darkgraycell{}\CIRCLE
    %\diagfil{0.68cm}{gray!25}{white}{\CIRCLE} 
    &  & \darkgraycell{footnote}\CIRCLE\tablefootnote[9]{}
    %\diagfil{0.68cm}{gray!25}{white}{\hspace{0.4em}\CIRCLE$^{\text{9}}$} 
    &  & \darkgraycell{}\CIRCLE
    %\diagfil{0.68cm}{gray!25}{white}{\CIRCLE}  
    &  &    \tn  \hhline{|~|*{13}{-}} %\cline{2-14}
    & Cryptographic postal voting     & &  &\graycell{}\CIRCLE  &  &  & \darkgraycell{}\CIRCLE
    %\diagfil{0.68cm}{gray!25}{white}{\CIRCLE}  
    &  & \darkgraycell{footnote}\CIRCLE\tablefootnote[9]{}
    %\diagfil{0.68cm}{gray!25}{white}{\hspace{0.40em}\CIRCLE$^{\text{9}}$} 
    &  & \darkgraycell{}\CIRCLE
    %\diagfil{0.68cm}{gray!25}{white}{\CIRCLE} 
    &  &    \tn \hhline{|~|*{13}{-}} %\cline{2-14}
    & Machine voting with paper trail   & &  &\graycell{}\CIRCLE  &  &  & \darkgraycell{}\CIRCLE
    %\diagfil{0.68cm}{gray!25}{white}{\CIRCLE}  
    &  & \darkgraycell{footnote}\CIRCLE\tablefootnote[9]{}
    %\diagfil{0.68cm}{gray!25}{white}{\hspace{0.40em}\CIRCLE$^{\text{9}}$} 
    &  &  & \darkgraycell{}\CIRCLE
    %\diagfil{0.68cm}{gray!25}{white}{\CIRCLE}  
    & \tn \hhline{|~|*{13}{-}} %\cline{2-14}
    & Machine voting, no paper trail   & &  &\graycell{}\CIRCLE  &  &  & \darkgraycell{}\CIRCLE
    %\diagfil{0.68cm}{gray!25}{white}{\CIRCLE}  
    &  & \darkgraycell{footnote}\CIRCLE\tablefootnote[9]{}
    %\diagfil{0.68cm}{gray!25}{white}{\hspace{0.40em}\CIRCLE$^{\text{9}}$}  
    &  &  & \darkgraycell{}\CIRCLE
    %\diagfil{0.68cm}{gray!25}{white}{\CIRCLE} 
    &\graycell{}\CIRCLE    \tn \hhline{|~|*{13}{-}} %\cline{2-14}
%    & I-voting with code voting   & &  &\CIRCLE  &  &  &  &\CIRCLE  &\CIRCLE  &\CIRCLE  &\CIRCLE  &  &\CIRCLE    \tn \cline{2-14}
    & I-voting   &\LEFTcircle\tablefootnote[6]{} &\CIRCLE\tablefootnote[14]{} &\graycell{}\CIRCLE  &  &
    \darkgraycell{footnote}\CIRCLE\tablefootnote[13]{}
    %\diagfil{0.68cm}{gray!25}{white}{\hspace{0.65em}\CIRCLE$^{\text{13}}$}  
    &  &  & \hspace{0.40em}\graycell{}\CIRCLE\tablefootnote[9]{} &\graycell{}\CIRCLE  &  &  &\graycell{}\LEFTcircle    \tn 
    \hlineB{2}

\multirow{9}{*}{\begin{tabular}[c]{@{}l@{}}\textbf{Verification}\tn \textbf{of eligibility}\end{tabular}}
    & Paper voting      &\LEFTcircle\tablefootnote[3]{} &  &\graycell{}\CIRCLE  &  &  & \darkgraycell{}\CIRCLE %\diagfil{0.68cm}{gray!25}{white}{\CIRCLE}  
    &  &  &  &  &  &    \tn  \hhline{|~|*{13}{-}} %\cline{2-14}
    & Cryptographic paper voting      &\CIRCLE\tablefootnote[3]{} &  &\graycell{}\CIRCLE  &  &  &
    \darkgraycell{}\CIRCLE
    %\diagfil{0.68cm}{gray!25}{white}{\CIRCLE}  
    &  &  &  &  &  &    \tn  \hhline{|~|*{13}{-}} %\cline{2-14}
    & Postal voting     &\CIRCLE\tablefootnote[3]{} &  &\graycell{}\CIRCLE  &  &  &
    \darkgraycell{}\CIRCLE
    %\diagfil{0.68cm}{gray!25}{white}{\CIRCLE}  
    &  &  &  &  &  &
    \darkgraycell{}\LEFTcircle
    %\diagfil{0.68cm}{gray!25}{white}{\LEFTcircle}   
    \tn  \hhline{|~|*{13}{-}} %\cline{2-14}
    & Cryptographic postal voting     &\CIRCLE\tablefootnote[3]{} &  &\graycell{}\CIRCLE  &  &  &
    \darkgraycell{}\CIRCLE
    %\diagfil{0.68cm}{gray!25}{white}{\CIRCLE}  
    &  &  &  &   &  &
    \darkgraycell{}\LEFTcircle
    %\diagfil{0.68cm}{gray!25}{white}{\LEFTcircle}  
    \tn  \hhline{|~|*{13}{-}} %\cline{2-14}
    & Machine voting   &\CIRCLE\tablefootnote[3]{} &  &\graycell{}\CIRCLE  &  &  &
    \darkgraycell{}\CIRCLE
    %\diagfil{0.68cm}{gray!25}{white}{\CIRCLE}  
    &  &  &  &  &  &    \tn  \hhline{|~|*{13}{-}} %\cline{2-14}
%    & I-voting with code voting   & &  &\CIRCLE  &\CIRCLE  &  &  &  &\CIRCLE  &\CIRCLE  &  &  &    \tn \cline{2-14}
    & I-voting  &\LEFTcircle\tablefootnote[6]{} &  &\graycell{}\CIRCLE  &\graycell{footnote}\CIRCLE\tablefootnote[12]{}  &  &  &  & \hspace{0.40em}\graycell{}\CIRCLE\tablefootnote[9]{}  & &  &  &\graycell{}\CIRCLE    \tn
    \hlineB{2}
    
\multirow{9}{*}{\begin{tabular}[c]{@{}l@{}}\textbf{Verification}\tn \textbf{of delivery}\end{tabular}}  
    & Paper voting      & &  &  &  &  &  &  &  &  &  &  &    \tn  \hhline{|~|*{13}{-}} %\cline{2-14}
    & Cryptographic paper voting      &\CIRCLE &  &  &  &  &  & \graycell{}\CIRCLE\tablefootnote[8]{}  &  &  &  &  &    \tn  \hhline{|~|*{13}{-}} %\cline{2-14}
    & Postal voting     & &  &  &  &  &
    \darkgraycell{}\CIRCLE
    %\diagfil{0.68cm}{gray!25}{white}{\CIRCLE} '
    &  & \darkgraycell{footnote}\CIRCLE\tablefootnote[9]{}
    %\diagfil{0.68cm}{gray!25}{white}{\hspace{0.40em}\CIRCLE$^{\text{9}}$}  
    &  & \darkgraycell{}\CIRCLE
    %\diagfil{0.68cm}{gray!25}{white}{\CIRCLE}  
    &  &  \tn  \hhline{|~|*{13}{-}} 
    & Cryptographic postal voting     &\CIRCLE &  &  &  &  &  &\graycell{}\CIRCLE\tablefootnote[8]{}  &  &  &
    \darkgraycell{}\CIRCLE
    %\diagfil{0.68cm}{gray!25}{white}{\CIRCLE} 
    &  &  \tn \hhline{|~|*{13}{-}} 
    & Machine voting with paper trail   &$\;\,$\LEFTcircle\tablefootnote[2]{} &   &  &  &  &   &  &  &  &  &  &  \tn  \hhline{|~|*{13}{-}} %\cline{2-14}
    & Machine voting, no paper trail   & &  &  &  &  &  &  &   &  &  &  & \darkgraycell{}\CIRCLE
    %\diagfil{0.68cm}{gray!25}{white}{\CIRCLE}  
    \tn  \hhline{|~|*{13}{-}}
%    & I-voting with code voting   &$\triangle$ &  &  &\XBox  &  &  &  &  &  &  &  &\CIRCLE   \tn  \cline{2-14}
    & I-voting (individual verifiability)   &\CIRCLE &\CIRCLE  &  &  &  &  &  &  &  &  &  &\graycell{}\LEFTcircle \tn  \hhline{|~|*{13}{-}} %\cline{2-14}
    & I-voting (universal verifiability)   &$\;\,$\LEFTcircle\tablefootnote[2] &$\;\,$\LEFTcircle\tablefootnote[2]&  & &  &  &  &  &  &  &  &\graycell{}\LEFTcircle  \tn  
    \hlineB{2}
    
\multirow{9}{*}{\begin{tabular}[c]{@{}l@{}}\textbf{Verification}\tn \textbf{of ballot box}\tn \textbf{integrity}\end{tabular}}  
    & Paper voting      & &   &  &   & 
    \darkgraycell{}\CIRCLE
    %\diagfil{0.68cm}{gray!25}{white}{\CIRCLE}   
    & \darkgraycell{}\CIRCLE
    %\diagfil{0.68cm}{gray!25}{white}{\CIRCLE}  
    &  & \darkgraycell{footnote}\CIRCLE\tablefootnote[9]{}
    %\diagfil{0.68cm}{gray!25}{white}{\hspace{0.40em}\CIRCLE$^{\text{9}}$}  
    &  &  &  &   \tn \hhline{|~|*{13}{-}}
    & Cryptographic paper voting      &  &   &  &  &  &  &  & \hspace{0.40em}\graycell{}\CIRCLE\tablefootnote[9]{} &  &  &  &   \tn \hhline{|~|*{13}{-}}
    & Postal voting     & &   &  &  & 
    \darkgraycell{}\CIRCLE
    %\diagfil{0.68cm}{gray!25}{white}{\CIRCLE}  
    & \darkgraycell{}\CIRCLE
    %\diagfil{0.68cm}{gray!25}{white}{\CIRCLE}  
    &  & \darkgraycell{footnote}\CIRCLE\tablefootnote[9]{}
    %\diagfil{0.68cm}{gray!25}{white}{\hspace{0.40em}\CIRCLE$^{\text{9}}$}  
    &  &  &  &   \tn \hhline{|~|*{13}{-}}
    & Cryptographic postal voting     &  &   &  &  &  &  &  & \hspace{0.40em}\graycell{}\CIRCLE\tablefootnote[9]{}  &  &  &  &   \tn \hhline{|~|*{13}{-}}
    & Machine voting with paper trail   & &  &  &  &
    \darkgraycell{}\CIRCLE
    %\diagfil{0.68cm}{gray!25}{white}{\CIRCLE}   
    &  &  & \darkgraycell{footnote}\CIRCLE\tablefootnote[9]{}
    %\diagfil{0.68cm}{gray!25}{white}{\hspace{0.40em}\CIRCLE$^{\text{9}}$}  
    &  &  &  &   \tn \hhline{|~|*{13}{-}}
    & Machine voting, no paper trail   & &  &  &  & 
    \darkgraycell{}\CIRCLE
    %\diagfil{0.68cm}{gray!25}{white}{\CIRCLE}   
    &  \darkgraycell{}\CIRCLE  &  &   &  &  &  &
    \darkgraycell{}\CIRCLE
    %\diagfil{0.68cm}{gray!25}{white}{\CIRCLE}  
    \tn  \hhline{|~|*{13}{-}} %\cline{2-14}
%    & I-voting with code voting   &$\triangle$ &  &  &\hexstar  &  &  &  &\CIRCLE  &  &  &  &\CIRCLE   \tn  \cline{2-14}
    & I-voting (individual verifiability)   &  &  &  &\graycell{}\CIRCLE  & \graycell{}\CIRCLE  &  &  &\hspace{0.40em}\graycell{}\CIRCLE\tablefootnote[9]{}  &  &  &  &\graycell{}\CIRCLE   \tn \hhline{|~|*{13}{-}}
    & I-voting (universal verifiability)   &   & &  &  &  &  &  & \hspace{0.40em}\graycell{}\CIRCLE\tablefootnote[9]{}   &  &  &  &  \tn  
    \hlineB{2}
    
\multirow{9}{*}{\begin{tabular}[c]{@{}l@{}}\textbf{Verification}\tn \textbf{of tally }\tn \textbf{integrity}\end{tabular}}  
    & Paper voting      & &  &  &  &   & 
    \darkgraycell{}\CIRCLE
    %\diagfil{0.68cm}{gray!25}{white}{\CIRCLE}  
    &  & \darkgraycell{footnote}\CIRCLE\tablefootnote[9]{}
    %\diagfil{0.68cm}{gray!25}{white}{\hspace{0.40em}\CIRCLE$^{\text{9}}$}  
    &  &  &  &   \tn \hhline{|~|*{13}{-}}
    & Cryptographic paper voting      &  &  &  &  &  &  &  & \hspace{0.40em}\graycell{}\CIRCLE$^{\text{9}}$  &  &  &  &   \tn \hhline{|~|*{13}{-}}
    & Postal voting    & &  &  &  &  & 
    \darkgraycell{}\CIRCLE
    %\diagfil{0.68cm}{gray!25}{white}{\CIRCLE} 
    &  & \darkgraycell{footnote}\CIRCLE\tablefootnote[9]{}
    %\diagfil{0.68cm}{gray!25}{white}{\hspace{0.40em}\CIRCLE$^{\text{9}}$}  
    &  &  &  &   \tn \hhline{|~|*{13}{-}}
    & Cryptographic postal voting     &  &  &  &  &  &  &  & \hspace{0.40em}\graycell{}\CIRCLE\tablefootnote[9]{}  &  &  &  &   \tn  \hhline{|~|*{13}{-}} %\cline{2-14}
    & Machine voting with paper trail   & &  &  &\graycell{}\CIRCLE  &   &
    \darkgraycell{}\CIRCLE
    %\diagfil{0.68cm}{gray!25}{white}{\CIRCLE}  
    &  & \darkgraycell{footnote}\CIRCLE\tablefootnote[9]{}
    %\diagfil{0.68cm}{gray!25}{white}{\hspace{0.40em}\CIRCLE$^{\text{9}}$}  
    &  & \darkgraycell{}\CIRCLE
    %\diagfil{0.68cm}{gray!25}{white}{\CIRCLE} 
    &  &   \tn  \hhline{|~|*{13}{-}} %\cline{2-14}
    & Machine voting, no paper trail   & &  &  &\graycell{}\CIRCLE  &    &
    \darkgraycell{}\CIRCLE
    %\diagfil{0.68cm}{gray!25}{white}{\CIRCLE} 
    &  & \darkgraycell{footnote}\CIRCLE\tablefootnote[9]{}
    %\diagfil{0.68cm}{gray!25}{white}{\hspace{0.40em}\CIRCLE$^{\text{9}}$}  
    &  & \darkgraycell{}\CIRCLE
    %\diagfil{0.68cm}{gray!25}{white}{\CIRCLE} 
    &  &\graycell{}\CIRCLE   \tn \hhline{|~|*{13}{-}} %\cline{2-14}
%    & I-voting with code voting  &$\triangle$  & &  &\hexstar  &  &  &  &\CIRCLE  &  &  &  &\CIRCLE   \tn \cline{2-14}
    & I-voting (individual verifiability)  & &  &  &\graycell{}\CIRCLE  & \graycell{}\CIRCLE  &  &  &\hspace{0.40em}\graycell{}\CIRCLE\tablefootnote[9]{}  &  &  &  &\graycell{}\CIRCLE   \tn \hhline{|~|*{13}{-}} %\cline{2-14}
    & I-voting (universal verifiability)  &  & &  &  &  &  &  & \hspace{0.40em}\graycell{}\CIRCLE\tablefootnote[9]{} &  &  &  &   \tn 
    \hlineB{2}

\multicolumn{14}{l}{dark gray background $=$ subset of votes is affected in case of attack} 
\tn
    
\multicolumn{14}{l}{white background on a filled cell $=$ single vote is affected in case of attack} 
\tn

\multicolumn{14}{l}{gray background $=$ all votes are affected in case of attack \hspace{3mm}  $\CIRCLE=$ is trusted \hspace{3mm}  $\LEFTcircle=$ is trusted under conditions
}

\end{tabular}
}
\end{minipage}
\begin{minipage}{0.48\textwidth}
\footnotetext [1]{Threshold number of parties must be trusted.}
\footnotetext[2]{Trusted to perform/enable verification if voting system allows to.}
\footnotetext[3]{If reliable IDs are not available, the voter is trusted to provide valid information.}
\footnotetext[4]{Filled in postal ballots have to be stored securely.}
\footnotetext[5]{Trusted, unless code voting is used.}
\footnotetext[6]{Only own voting credentials may be used, they have to be secured.}
\footnotetext[7]{Recording devices placed by infrastructure provider.}
\end{minipage}
~
\begin{minipage}{0.48\textwidth}
\footnotetext[8]{If codes are misprinted, ballots can not be verified.}
\footnotetext[9]{Distributed trust, assumed to behave honestly.}
\footnotetext[10]{Possible to check whether voter abstained.}
\footnotetext[11]{For pre-printed ballots we need to trust printing house and software\tn\hspace{4cm} vendor. For print-on-demand, we need to trust software vendor.}
\footnotetext[12]{Trusted to ensure that only ballot from eligible voters are processed.}
\footnotetext[13]{Servers may be taken offline.}
\footnotetext[14]{Trusted not to prevent voter from voting.}
\end{minipage}

\end{table*}

\end{document}